\documentstyle[aps]{revtex}

\def \f{\frac}

\let \pp=\partial

\newcommand{\be}{\begin{equation}}
\newcommand{\ee}{\end{equation}}
\newcommand{\bea}{\begin{eqnarray}}
\newcommand{\eea}{\end{eqnarray}}
\newcommand{\bean}{\begin{eqnarray*}}
\newcommand{\eean}{\end{eqnarray*}}

\def \ni{\noindent}

\def \mn{\medskip
\ni}
\def \tr{\textrm}
\newcommand{\N}{\mathrm{I\hspace{-.37ex}N}}
\newcommand{\R}{\mathrm{I\hspace{-.37ex}R}}

\def \w{\wedge}
\newcommand{\Ref}[1]{(\ref{#1})}
\def \tl{\tilde}

\begin{document}

\title{\Large \bf Barrett-Crane spin foam model \\ from generalized
BF-type action for gravity}

\author{\bf Richard E. Livine} 

\address{Centre de Physique Th\'eorique - CNRS, Case 907, Luminy,
F-13288 Marseille, France \\
and \\
The Blackett Laboratory, Imperial College of Science, Technology
and Medicine, \\
South Kensington, London SW7 2BZ, UK \\ e.livine@ic.ac.uk}

\author{\bf Daniele Oriti}

\address{Department of Applied Mathematics and Theoretical Physics, \\
Centre for Mathematical Sciences, \\ University of Cambridge, \\
Wilberforce Road, Cambridge CB3 0WA, UK \\ d.oriti@damtp.cam.ac.uk}

\date{April 13, 2001}
\maketitle

\begin{abstract}
We study a generalized action for gravity as a constrained BF theory, and
its relationship with the Plebanski action. We analyse the discretization
of the constraints and the spin foam quantization of the theory, showing
that it leads naturally to the Barrett-Crane spin foam model for quantum
gravity. Our analysis holds true in
both the Euclidean and Lorentzian formulation.  
\end{abstract}

\section{Introduction}

\ni
Spin foam models \cite{baez} \cite{baez2} are an attempt to formulate a
non-perturbative and background-independent theory of quantum gravity, and
can be seen as a new way to construct a sum-over-geometries out of algebra
and combinatorics, with the geometrical quantities emerging in the
semiclassical limit only. With a striking convergence of results and
ideas, spin foam models emerged \cite{Rov0} also in contexts as different
as canonical loop quantum gravity \cite{mike&carlo} and discrete
topological field theory \cite{T-V}. The most studied and developed of
these is the Barrett-Crane model for both Euclidean \cite{BC} and
Lorentzian \cite{BC2} quantum gravity.
It was first obtained as a quantization of
simplicial geometry, using the methods of category theory and geometric
quantization.
It was then re-derived using a field theory over a group
manifold
\cite{dfkr}\cite{rr}\cite{rr2}\cite{alej}\cite{alej2}\cite{alej3}, or
generalized matrix model.
And finally, it was shown to arise from a discretization
and quantization of BF theory \cite{daniele}. 

\mn
Recently, a new action for gravity as a constrained BF theory was proposed
\cite{CMPR} and it seemed that a discretization and
spin foam quantization of
it would lead to a model necessarily different from the Barrett-Crane
one. Also, a natural outcome would be a one-parameter ambiguity in the
corresponding spin foam model, related to the Immirzi parameter of
loop quantum gravity \cite{immirzi}\cite{tom&carlo}, and this
gives an additional reason to study the spin foam quantization of
the generalized action since it could help understand the link
between the current spin foam models and loop quantum gravity.

\mn
We show here that a careful discretization of the new
form of the constraints, an analysis of the field content of the theory,
at the classical level, and a spin foam quantization taking all this into
account, lead naturally to the Barrett-Crane model as a quantum theory
corresponding to that action. No one-parameter ambiguity arises in the
spin foam model and in the quantum geometry described by it. This suggests that the Barrett-Crane model is more
universal than
at first thought and that its continum limit may be described by several
different lagrangians.
Finally,
in section \ref{area},
we also briefly discuss the issue of the area spectrum in the spin foam 
framework. 

\section{The Barrett-Crane model} \label{BC}
First of all, we recall the basic elements of the Barrett-Crane model
\cite{BC}\cite{BC2}. Consider a geometric 4-simplex, i.e. the convex hull
of 5 points in $\R^{4}$. It determines and is uniquely
characterized (up to parallel translation and inversion through the
origin) by a set of 10 bivectors
$b_{i}\in\R^{4}\wedge\R^{4}$, assigned one to each of its
triangles, satisfying the following properties: 
\begin{itemize}
\item a different orientation of the triangle corresponds to a bivector
with opposite sign;
\item the bivectors assigned to the triangles are simple, i.e. given by a
wedge product of two vectors;
\item two bivectors corresponding to two triangles sharing an edge add to
a simple bivector; 
\item the (oriented) sum of the 4 bivectors assigned to the 4 triangles of
a tetrahedron is zero;
\item the six bivectors corresponding to six triangles sharing a vertex
are linearly independent;
\item given 3 triangles meeting at a vertex of a tetrahedron,
the volume spanned by the 3 corresponding bivectors must be postive
i.e the bivectors (considered as operators by means of the metric)
satisfy: $Tr\,b_{1}[b_{2},b_{3}]>0$.
\end{itemize}   

\ni
For quantum 4-simplices, we deal with possibly degenerate 4-simplices,
so we drop the linear independence (condition 5) and allow for zero
volume in the last condition ($\ge 0$ instead $>0$).
Now we can proceed to the quantization. We associate to each triangle an
element of the Lie algebra of the local gauge group  ($so(4)$ in the
Euclidean case, and $so(3,1)$ in the lorentzian) using the isomorphism
between bivectors and Lie algebra elements, and then turn them into
operators choosing a representation, so that we obtain bivector operators
acting on the Hilbert space given by the representation space chosen. To
each tetrahedron is then associated an element of the tensor product of
the 4 Hilbert spaces associated to its faces. To characterize completely
the quantum geometry of the 4-simplex the chosen representations have to
satisfy a quantum analog of the conditions above:
\begin{itemize}
\item the representations corresponding to different orientations of the
same trangle are dual to each other;
\item the representations used are simple representations, i.e.
characterized by a vanishing second casimir of the algebra;
\item given two faces of a tetrahedron, we can decompose the corresponding
pair of representations into its Clebsch-Gordon series; then the tensor
for the tetrahedron have to decompose into summands given by simple
representations only;
\item the tensor for the tetrahedron is invariant under the local gauge
group.
\end{itemize}
In the Euclidean case a bivector is simple when its selfdual and
anti-selfdual parts have the same magnitude. The splitting into selfdual
and anti-selfdual parts corresponds to the splitting of the $Spin(4)$
algebra (universal covering of $so(4)$)
into a sum of two $su(2)$ algebra, so that the irreducible
representations are given by pairs of spins $(j^+,j^-)$.
The simplicity of these representations corresponds to the
vanishing of the second casimir, $\epsilon_{IJKL}J^{IJ}J^{KL}$
given in the canonical
basis,  $C_{2}=j^+(j^++1)-j^-(j^-+1)$. This implies that $j^+=j^-$ so
that the representations to be used are of the type $(j,j)$.
In the Lorentzian case,
the same splitting is possible only through complexification of
the fields. Nevertheless, we can work with the principal series
of irreducible unitary representations of $so(3,1)$
\cite{BC2}\cite{alej2}\cite{alej3}. They are labelled by
a half-integer number $j$ and a positive real number
$\rho$. The second casimir of $so(3,1)$ is given by $C_{2}=\f{1}{2}j\rho$
so
that simplicity implies that we restrict ourselves
to using representations corresponding
to the two series $(j,0)$ and $(0,\rho)$.  

\mn
With these conditions, an amplitude for a 4-simplex can be obtained
\cite{BC}\cite{BC2} (as the evaluation of a spin network), and
one can characterize completely the 4-geometry at the quantum level
\cite{baez2}\cite{BB}, and construct a complete spin foam model for a
triangulated manifold
\cite{baez2}\cite{dfkr}\cite{alej}\cite{alej2}\cite{alej3}\cite{daniele}.
This model, both in its Euclidean version and in its Lorentzian
formulation, is shown to be finite given a fixed triangulation
(we sum over all possible colorings of the faces) 
and therefore the spin foam model is well-defined \cite{finite}
\cite{finite2}\cite{finite3}.
We don't give here the details of this construction for which we refer to
the literature. Instead we think is useful for our pourposes to describe
how the association of bivectors to the triangles of a 4-simplex can be
naturally made, in a gravitational context. Suppose we have a tetrad field
$e: TM_{p}\rightarrow\R^{4}$, where $M$ is our spacetime manifold,
so that $e\wedge e:\wedge^{2}( TM_{p})\rightarrow\wedge^{2}(
\R^{4})$, mapping any wedge product of vectors $u_{1}\wedge u_{2}$
into a bivector $e(u_{1})\wedge e(u_{2})$. Then the bivector $b_{T}$
associated to the triangle $T$ is naturally given by: $b_{T}=\int_{T}
e\wedge e=e(u)\wedge e(v)$, where $u$ and $v$ are the vectors
corresponding to two edges of the triangle.

\mn
The Barrett-Crane model was argued to be a quantization of the Plebanski
action\cite{Pleb} for gravity in \cite{dpf}, and this conclusion is also
supported by the results of \cite{daniele}. The Plebanski action is a a
BF-type action, in the sense that it gives gravity as a constrained BF
theory, with quadratic constraints on the B field (we note that such
a formulation of general relativity has been generalised to any dimension
\cite{fkp}). More precisely the action is given by:
\be 
S\,=\,S(\omega,B,\phi)\,=\,\int_{\mathcal{M}}\left[
B^{IJ}\,\wedge\,F_{IJ}(\omega)\,-\frac{1}{2}\phi_{IJKL}\,B^{KL}\,\wedge\,B^{IJ}\right]
\ee
where $\omega$ is a connection 1-form valued in $so(4)$ ($so(3,1)$),
$\omega=\omega_{a}^{IJ}J_{IJ}dx^{a}$, $J_{IJ}$ are the generators of
$so(4)$ ($so(3,1)$, $F=d\omega$ is the corresponding two-form curvature,
$B$ is a 2-form also valued in $so(4)$ ($so(3,1)$),
$B=B_{ab}^{IJ}J_{IJ}dx^{a}\wedge dx^{b}$, and $\phi_{IJKL}$ is a
Lagrange multiplier satisfying $\phi_{IJKL}\epsilon^{IJKL}=0$. Here and in
the following $a,b,..$ are spacetime indices and $I,J,K,...$ are internal
indices. The equations of motion are:
\be
d B\,+\,[\omega,B]=0 \;\;\;\;\;\;\;\;
F^{IJ}(\omega)\,=\,\phi^{IJKL}B_{KL}\;\;\;\;\;\;\;\; 
B^{IJ}\,\wedge\,B^{KL}\,=\,e\,\epsilon^{IJKL} \label{eq:constrB}
\ee
where $e=\frac{1}{4!}\epsilon_{IJKL}B^{IJ}\wedge B^{KL}$.
When $e\ne0$, the constraint \Ref{eq:constrB}
is equivalent to
$\epsilon_{IJKL}B^{IJ}_{ab}B^{KL}_{cd}=\epsilon_{abcd}e$
\cite{dpf}, which implies that
$\epsilon_{IJKL}B^{IJ}_{ab}B^{KLab}=0$
i.e. $B_{ab}$ is a simple bivector.
Moreover, \Ref{eq:constrB}
is satisfied if and only if there exists a real tetrad field 
$e^{I}=e^{I}_{a}dx^{a}$ so that one of the following equations holds:
\bea &I&\;\;\;\;\;\;\;\;\;\;B^{IJ}\,=\,\pm\,e^{I}\,\wedge\,e^{J} \\ 
&II&\;\;\;\;\;\;\;\;\;\;B^{IJ}\,=\,\pm\,\frac{1}{2}\,\epsilon^{IJ}\,_{KL}e^{K}\,\wedge\,e^{L}
.  
\eea
Restricting the field B to be always in the sector $II_{+}$ (which is
always possible classically), the action becomes:
\be 
S\,=\,\int_{\mathcal{M}}\,\epsilon_{IJKL}\,e^{I}\,\wedge\,e^{J}\,\wedge\,F^{KL}
\ee
which is the action for general relativity in the first order Palatini
formalism. Then a discretization of the constraints ~\ref{eq:constrB},
giving a bivector field from the B field for each triangle, shows
\cite{dpf} that they correspond exactly to the conditions given above for
the bivectors characterizing a 4-simplex, and this leads to the conclusion
that the Barrett-Crane model gives the quantization of the Plebanski
action.

\mn
More precisely, going from the field theory to the Barrett-Crane model
is achieved in 3 main steps. First, it is the {\it discretization} of
the two-form into bivectors associated to each face to the triangulation.
Then, we translate bivectors as elements of $so(4)^*$ or $so(3,1)^*$,
as described in \cite{BB}, using the function

\begin{equation}
\begin{array}{ccccc}
\theta &:& \Lambda^2 \R^4  & \rightarrow & so(4)^* \tr{ or } so(3,1)^* \\
 & & e \w f & \rightarrow & \theta(e\w f) \, (l) \rightarrow \eta(l e,f)
\end{array}
\end{equation}

\ni
where $\eta$ is the Euclidean or Lorentzian metric. Less formally, it is
the step we call {\it correspondence} between bivectors $B$ and elements
of the Lie algebra $J$. The last step is the {\it quantization} itself,
using
techniques from geometric quantization. This gives representation
labels to the faces of the triangulation and gives the Barrett-Crane model
(to some normalisation factors).

\mn
In fact, there is an ambiguity at the level of the correspondence. We can
also choose to use the isomorphism $\theta \tr{o} *$ where $*$ is the
Hodge operator.
This leads to the so-called
flipped Poisson bracket, and it is indeed the right thing to do.
In the Euclidean case, this leads us to only ``real'' tetrahedra, whose
faces
are given by the bivectors and not the Hodge dual of the bivectors. This
amounts to selecting the sector $II$ of the theory which is the sector
we want \cite{BB}. In the Lorentzian case, such a check on the tetrahedra
hasn't been done yet, however the use of the flipped correspondence has a
nice consequence: it changes the sign of the area to being given by $-C_1$
instead of $C_1$ so that the discrete series of representations $(n,0)$
truly correspond to time-like faces and the continuous series $(0,\rho)$
to space-like faces, as implied by the algebraic
properties of these representations \cite{BC2}.

\ni
Nevertheless, we can generalize this correspondence. We have a familly
of such isomorphisms given by $\theta \tr{o} (\alpha* +\beta)$. For
$\alpha$
and $\beta$ different from 0, it wouldn't give anything interesting when
dealing with the Barrett-Crane conditions. However, it is this
generalized correspondence we are going to use to deal with the
generalized BF-type action.
And, at the end, we will find again the same Barrett-Crane model.

\section{Generalized BF-type action for gravity} \label{CMPR}

\label{debut}

\ni
In \cite{CMPR} the following BF-type action was proposed for general
relativity:

\be
S=\int B^{IJ}\w F_{IJ} -\f{1}{2}\phi_{IJKL}B^{IJ}\w B^{KL}+\mu H
\label{action}
\ee

\ni
where $H=a_1\phi_{IJ}\,^{IJ}+a_2\phi_{IJKL}\epsilon^{IJKL}$, where $a_{1}$
and $a_{2}$ are arbitrary constants.
$B$ is a 2\_form and $F$ is the curvature associated to the connection
$\omega$.
$\phi$ (spacetime scalar) and $\mu$ (spacetime 4-form) are Lagrange
multipliers, with $\phi$ having
the symmetries $\phi_{IJKL}=-\phi_{JIKL}=-\phi_{IJLK}=\phi_{KLIJ}$.
$\phi$ enforces the constraints on the $B$ field, while $\mu$ enforces
the condition $H(\phi)=0$ on $\phi$.
The $*$ operator acts on internal indices so that
$*B_{IJ}=1/2\,\epsilon_{IJKL}B^{KL}$ and $*^2=\epsilon$, with $\epsilon=1$
in the Euclidean case and $\epsilon=-1$ in the Lorentzian.

\mn
Before going on, we would like to point out that this is indeed the most
general action that can be built out of BF theory with a quadratic
constraint on the B field. This in turn means that the scalar constraint
$H=0$ is the most general one that can be constructed with a $\phi$ with
the given symmetry properties. This can be proven very easily. First of
all, we note that the two scalars $\phi_{IJ}\,^{IJ}$ and
$\phi_{IJKL}\epsilon^{IJKL}$ are linearly independent as it is immediate
to verify, so that what we have to prove is just that the space of scalars
made out of $\phi$ is 2-dimensional. There is an easy way to see this in
the Euclidean case: $\phi$ is a tensor in the 4-dimensional representation
of $so(4)$, which, using the splitting $so(4)\simeq su(2)\oplus su(2)$,
can be thought as given by a sum of two 2-dimensional representations of
$su(2)$, that in turn can be decomposed into a sum of a 3-dimensional (and
symmetric) one and a 1-dimensional (and antisymmetric) one. Now we have
just to compute the tensor product of 4 such representations, paying
attention in keeping only the terms with the desired symmetry properties,
to see that we can have two and only two resulting singlets.   

\mn
The equations of motion for $\omega$ and $B$ are the same as those coming
from the Plebanski action, but the constraints on the field B now are:

\be
B^{IJ}\w B^{KL} =\f{1}{6} (B^{MN}\w B_{MN}) \eta^{[I |K|} \eta^{J]L}
+\f{\epsilon}{12}(B^{MN}\w *B_{MN}) \epsilon^{IJKL}
\label{whole}
\ee

\be
2a_2 B^{IJ}\w B_{IJ} -\epsilon a_1 B^{IJ}\w *B_{IJ}=0
\label{simple}
\ee

\ni
The solution of these constraints \cite{prieto}, for non-degenerate $B$
($B^{IJ}\w *B_{IJ}\neq 0$), is:

\be
B^{IJ}=\alpha *(e^I \w e^J) + \beta\, e^I \w e^J
\label{B}
\ee

\ni
with:

\be
\f{a_2}{a_1}=\f{\alpha^2+\epsilon\beta^2}{4\alpha\beta}
\label{a1a2}
\ee

\ni
Inserting this solution into \Ref{action}, we get:

\be
S=\alpha \int  *(e^I \w e^J)\w F_{IJ}  +
\beta \int e^I \w e^J \w F_{IJ} \label{Holst}
\ee

\ni
so that there is a coupling between the geometric sector given by $*(e\w
e)$
(general relativity)
and the non-geometric one given by $e\w e$.
Nevertheless, we note that the second
term vanishes on shell so that the equations of motion ignore the
non-geometric part and are still given by the Einstein equations.

\mn
In the usually studied case $a_1=0$, \cite{dpf}\cite{fkp},
we are back to the Plebanski action, the sectors of
solutions being given by $\alpha=0$ and $\beta=0$ so that
we have either the general relativity sector or the non-geometric sector
$e\w e$.

\ni
In the particular case $a_2=0$, in the Euclidean case, the only
solution to \Ref{a1a2} is $\alpha=\beta=0$
so that only degenerate tetrads are going to contribute.
On the other hand, in the Lorentzian case we have instead
$\alpha=\pm\beta$.

\mn
Looking at \Ref{a1a2}, once we have chosen a couple $(\alpha,\beta)$,
we see that we have four posible sectors as with the Plebanski action
\cite{dpf}. In the Euclidean case, we can exchange $\alpha$ and $\beta$.
Under this transformation, the B field gets changed into
its Hodge dual, so we can trace back this symmetry to the fact that we can
use both $B$ and $*B$ as field variables in our original action, without
any change in the physical content of the theory. We
can also change  $B\rightarrow -B$ without affecting the physics of our
model.
This gives us the following four sectors:

\be
(\alpha,\beta) \quad (-\alpha,-\beta) \quad (\beta,\alpha) \quad
(-\beta,-\alpha)
\ee

\ni
In the Lorentzian case, the same $*$-symmetry brings us the following four
sectors:

\be
(\alpha,\beta) \quad (\beta,-\alpha) \quad
(-\alpha,-\beta) \quad (-\beta,\alpha) 
\ee

\mn
The canonical analysis of the action \Ref{Holst} was performed in
\cite{holst}, leading to the presence of the Immirzi parameter of
loop quantum gravity given by $\gamma=\alpha/\beta$ and related to $a_1$
and $a_2$ by:

\be
\f{a_2}{a_1}=\f{1}{4}\left(\gamma +\f{\epsilon}{\gamma}\right)
\ee

\ni
We can notice that we have two sectors
in our theory with different Immirzi parameters: $\gamma$
and $\epsilon/\gamma$, corresponding to a symmetry exchanging $\alpha$ and
$\epsilon\beta$.

\mn
The full symmetry group of the theory is then $Diff(M)\times
SO(4)\times Z_{2}\times Z_{2}$, with $SO(4)$ replaced by
$SO(3,1)$ in the Lorentzian case.  The $Z_{2}\times Z_{2}$
comes from the existence of the four sectors of solutions and
is responsible for their interferences at the quantum level.

\mn 
We want to study the relationship between this new action and the usual
Plebanski action, and its spin foam quantization, in order to understand
if and in which cases the corresponding spin foam model at the quantum
level is still given by the Barrett-Crane one.
To this aim it is important to note that the constraints \Ref{whole} and
\Ref{simple}
can be recast in an equivalent
form (see in appendix for more details)
leading to the same set of solutions,
for $a_2\neq 0$, $B$ non-degenerate, and
$\left(\f{a_1}{2a_2}\right)^2\neq\epsilon$
(this excludes the purely
selfdual and the purely anti-selfdual cases).
The situation is analogous to
that analyzed in \cite{dpf} for the Plebanski action. The
new constraint is

\be
\left(\epsilon_{IJMN}-\f{a_1}{a_2}\eta_{[I\mid M\mid}\eta_{J]N}
\right)
B^{MN}_{cd}B^{IJ}_{ab}=
e\,\epsilon_{abcd}
\left(1-
\epsilon\left(\f{a_1}{2a_2}\right)^2
\right)
\label{simpl2}
\ee

\ni
where

\be
e=\f{1}{4!}\epsilon_{IJKL}B^{IJ}\w B^{KL}
\ee

\ni
This constraint can then be discretised to give the simplicity constraint
and the intersection constaint leading to the Barrett-Crane model, as we
will see in section \Ref{class}. But we can already
notice that in the case $(ab)=(cd)$, \Ref{simple2} gives an
equivalent to \Ref{simple}:

\be
2a_2\f{1}{2}\epsilon_{IJKL}B^{IJ}_{ab}B^{KLab}
-\,a_1B^{IJ}_{ab}B^{ab}_{IJ}=0
\label{simple2}
\ee

\ni
In the following, we mainly use this second form of the constraints,
discussing the discretization and possible
spin foam quantization of the first one
in section \Ref{reis}.

\mn
Looking at the constraints on $B$ it is apparent that they are not anymore
just simplicity constraints like in the Plebanski case, so that a
direct
discretization of them for $B$ would not give the Barrett-Crane
constraints. We can see this by translating the constraints into
a condition on the Casimirs of $so(4)$ (or $so(3,1)$), using the
isomorphism between bivectors and Lie algebra.
We can naively replace
$B^{IJ}_{ab}$ with the canonical
generators $J^{IJ}$ of $so(4)$ (or $so(3,1)$), giving the correspondence:

\bea
B^{IJ}_{ab}B^{ab}_{IJ} &
\rightarrow & J^{IJ}J_{IJ}=2C_1\\
\f{1}{2}\epsilon_{IJKL}B^{IJ}_{ab}B^{KLab} &
\rightarrow & \f{1}{2}\epsilon_{IJKL}J^{IJ}J^{KL}
=2C_2
\eea

\ni
Using this, \Ref{simple2} gets transformed into:

\be
2a_2 C_2 - \, a_1 C_1 =0
\label{mixed}
\ee

\ni
or equivalently:

\be
2\alpha\beta C_1 = (\alpha^2+\beta^2)C_2
\ee

\ni
Then we would conclude as suggested in \cite{CMPR}
that we should use non-simple representations
in our spin foam model.
Actually the situation would be even worse than
this, since it happens that in general (for arbitrary values of $a_1$ and
$a_2$) no spin foam model can be constructed using only representations
satisfying \Ref{mixed} in the Euclidean case. 
More precisely,
using the splitting of the algebra $so(4)\simeq su(2)_+\oplus su(2)_-$,
the two Casimirs are:

\bea
C_1&=&j^+(j^++1)+j^-(j^-+1) \nonumber \\
C_2&=&j^+(j^++1)-j^-(j^-+1)
\eea

\ni
Apart from the case $a_1=0$ which gives us the Barrett-Crane simplicity
constraint $C_2=0$, the equation
\Ref{mixed} have an infinite number of solutions only when
$2a_2=\pm a_1$, in which cases we get representations of the form
$(j^+,0)$ (or $(0,j^-)$). This could be expected since the constraint
\Ref{simple} with this particular value of the parameters implies that the
$B$ field is selfdual or anti-selfdual when non-degenerate.
In the other cases, \Ref{mixed}
can be written as $j^+(j^++1)=\lambda j^-(j^-+1)$, with
$2a_2=(1+\lambda)/(1-\lambda)a_1$ and in general has no solutions.
For particular values of $\lambda$, it can have one and only one solution.
This would lead us to an ill-defined spin foam model using only one
representation $(j^+_0,j^-_0)$.
However, using the framework set up in \cite{mike3},
it can be easily proven that
it is not possible to construct an intertwiner for
a spin network built out of a single representation
(a single representation is not stable under
change of tree expansion for the vertex) so that no spin foam model
can be created.

\ni
In the Lorentzian case, the situation is more complex. The representations
are labelled by couples $(j\in\N/2,\rho\ge0)$ and the two Casimirs
are \cite{alej2}\cite{ruhl}:

\bea
C_1&=&j^2-\rho^2-1 \nonumber \\
C_2&=&\f{1}{2}j\rho
\eea

\ni
The equation \Ref{mixed} now reads:

\be
\rho^2+\f{a_2}{a_1}j\rho-j^2+1=0
\ee

\ni
and admits the following solutions if $a_1\ne 0$:

\be
\rho=\f{1}{2}\left[-\f{a_2}{a_1}j
\pm \sqrt{\left(\f{a_2}{a_1}\right)^2 j^2 +4(j^2-1)}\right]
\ee

\ni
So we always have some solutions to the mixed simplicity condition
\Ref{mixed}. However, instead of having a discrete series of
representations $(n,0)$ and a continuous series $(0,\rho)$, which are said
to correspond
to space-like and time-like degrees of freedom, as in the simple case
$a_1=0$, we end up with a (couple of) discrete series representations with
no direct interpretation. We think it is not possible to construct a
consistent vertex using them (it is also hard to imagine how
to construct a field theory over a group manifold formulation
of such a theory whereas, in the case of simple representations,
the construction was rather straightforward \cite{alej2}\cite{alej3}
adapting the Euclidean case to the Lorentzian case).
But this should be investigated.

\mn
As we will show, on the contrary, a careful analysis of the field content
of the theory, and of the correspondence between Lie algebra elements and
bivectors shows that not only a spin foam quantization is possible, but
that the resulting spin foam model should again be based on the
Barrett-Crane quantum constraints on the representations.

\section{Field content and relationship with the Plebanski action}
\label{class}

We have seen that the $B$ field, even if subject to constraints which are
more complicated than the Plebanski constraints, is forced by them to be
in 1-1 correspondence with the 2-form built out of the tetrad field,
$*(e\w e)$, which in turn is to be considered the truly physical field of
interest, since it gives the geometry of the manifold through the Einstein
equations. To put it in another way, we can argue that the physical
content of the theory, expressed by the Einstein equations, is independent
of the fields we use to derive it. In a discretized context, in
particular, we know that the geometry of the manifold is captured by
bivectors associated to 2-dimensional simplices, and constrained to be
simple. Consequently we would expect that in this context we should be
able to put the $B$ field in 1-1 correspondence with another 2-form field,
say $E$, then discretized to give a bivector for each triangle, in such a
way that the mixed constraints \Ref{simple2} would imply the simplicity of
this new bivector and the other Barrett-Crane constraints. If this
happens, then it would also mean that the action is equivalent to the
Plebanski action in terms of this new 2-form field, at least for what
concerns the constraints. This is exactly the case, as we are going to
prove.

\mn
The correspondence between $B$ and $E$ is actually suggested by the form
of the solution \Ref{B}. We take

\be
B_{ab}^{IJ}\,=\,\left( \alpha\,I\,+\,\epsilon\,\beta\,*\right)E^{IJ}_{ab}
 \label{transf} 
\ee

\ni
This is an invertible transformation, so that it really gives a 1-1
correspondence, if and only if $\alpha^{2}-\epsilon\beta^{2}\neq 0$, which
we will assume to be the case in the following.

\mn
Formally, we can do such a change of variable directly on the action
and express the action itself in terms of the field $E$, making
apparent that the constraints are just the Plebanski constraints:

\be
\left\{
\begin{array}{ccc}
B^{IJ}& =&\alpha E^{IJ}+\epsilon\beta *E^{IJ} \\
\tl{\phi}_{IJKL}&=&
(\alpha+\epsilon\beta\f{1}{2}\epsilon_{IJ}\,^{AB})\phi_{ABCD}
(\alpha+\epsilon\beta\f{1}{2}\epsilon_{KL}\,^{CD})
\end{array}
\right.
\label{change}
\ee

\ni
After this change, the action \Ref{action} becomes:

\be
S\,=\,
\f{1}{|\alpha^2-\epsilon\beta^2|^3}\,
\int\,(\alpha E^{IJ}+\epsilon\beta*E^{IJ})\wedge F_{IJ}\,
-\,\f{1}{2}\tl{\phi}_{IJKL}E^{IJ}\wedge E^{KL}
\,+\,\mu\epsilon^{IJKL}\tl{\phi}_{IJKL}
\label{after}
\ee

\ni
so that we have the Plebanski constraints on the $E$ field and we can
derive
directly from this expression the Holst action \Ref{Holst}. In the
Euclidean
case, decomposing  into selfdual and anti-selfdual components can be
quite useful to understand the structure of the theory. This is done
in the appendix and shows that the previous change of variable is simply
a rescaling of the selfdual and anti-selfdual parts of the $B$
field.

\mn
Let's now look at the discretization of the constraints. We will follow
the
same procedure as in \cite{dpf}. 
Using the 2-form $B$, a bivector can be associated to any 2-surface $S$ in
our manifold by integrating the 2-form over the surface:

\be
B^{IJ}(S)\,=\,\int_{S}B^{IJ}\,=\,\int_{S}\left(
\alpha\,I\,+\,\epsilon\,\beta\,*\right)E^{IJ}\,
=\,\alpha\,E^{IJ}(S)\,+\,\epsilon\,\beta\,*E^{IJ}(S)
\label{dis}.
\ee

\ni
Note that this automatically implements the first of the Barrett-Crane
constraints for the bivectors $E(S)$ and $B(S)$ (a change of orientation
of the surface $S$ will change the sign of the bivectors).

\ni
We now take a triangulation of our manifold, such that $B$ is constant
inside each 4-simplex, i.e. $dB=0$, and we associate a bivector to each
triangle of the triangulation using the procedure above.
This is equivalent to supoosing $E$ constant in the 4\_simplex
since the map \Ref{transf} is invertible.
Then we can use
Stokes' theorem to prove that the sum of all the bivectors $E$ associated
to the 4 faces $t$ of a tetrahedron $T$ is zero:


\be
0 \,=\, \int_{T}dE \,=\, \int_{\partial T}E
\,=\,\int_{t_1}E\,+\int_{t_2}E\,+\int_{t_3}E\,+\int_{t_4}E
\,=\, E(t_1)+E(t_2)+E(t_3)+E(t_4)
\ee

\ni
meaning that the bivectors $E$ satisfy the fourth Barrett-Crane
constraints (closure constraint).

\ni
Let's consider the constraints in the form \Ref{simpl2}.
Using \Ref{transf}, we obtain a much simpler
constraint on the $E$ field:

\be
\left(\alpha^{2} \,+\, \epsilon\,\beta^{2}\right) \,
\epsilon_{IJKL}\,E^{IJ}_{ab}\,E^{KL}_{cd}
\,=\,
\,e\,\epsilon_{abcd}
\label{discon}
\ee

\ni
Now,
$e=\f{1}{4!}\epsilon_{IJKL}B^{IJ}\w B^{KL}=
\f{1}{4!}(\alpha^2+\epsilon\beta^2)\epsilon_{IJKL}E^{IJ}\w E^{KL}$
is a sensible volume
element, since we assumed that the $B$ field is non-degenerate.

After imposing the equation of motion,
it appears that it is also the ``right'' geometric
volume element, i.e. the one constructed out of the tetrad field.
More precisely, equation \Ref{discon}
implies the simplicity of the field $E$. Thus there exist a tetrad field
such that $E^{IJ}=\pm e^I\w e^J$ or $E^{IJ}=\pm*(e^I\w e^J)$ and
this tetrad field is the one defining the metric after imposing the
Eisntein equation.
The scalar $e$ is then proportional to
$\epsilon_{IJKL}e^{I}\wedge e^{J}\wedge e^{K}\wedge
e^{L}=det(e)$. Consequently, up to a factor, the 4-volume  
spanned by two faces $t$ and $t'$ of a 4-simplex is given by:

\be
V(t,t')\,=\,\int_{x\in t\,;\,y\in t'}\,e\,\epsilon_{abcd}\,dx^{a}\w
dx^{b}\w dy^{c}\w dy^{d}
\ee

\ni
Then, integrating equation \Ref{discon} gives directly:

\be
\epsilon_{IJKL}\,E^{IJ}(t)\,E^{KL}(t')
=\f{1}{\left( \alpha^{2}\,+\,\epsilon\,\beta^{2}\right)}\,V(t,t')
\label{cons}
\ee

\ni
Considering only one triangle $t$:

\be
\epsilon_{IJKL}\,E^{IJ}(t)\,E^{KL}(t)=0
\ee

\ni
so that for any of the bivectors $E$ the selfdual part has the same
magnitude of the anti-selfdual part,
So that the $E(t)$ are simple bivectors.
This corresponds to the second of the Barrett-Crane constraints
(simplicity constraint).

\ni
For two triangles sharing an edge, we similarly have:

\be
\epsilon_{IJKL}\,E^{IJ}(t)\,E^{KL}(t')=0
\ee

\ni
This can be rewritten as:

\be
\epsilon_{IJKL}(E^{IJ}(t)+E^{IJ}(t'))(E^{KL}(t)+E^{KL}(t'))
-\epsilon_{IJKL}\,E^{IJ}(t)\,E^{KL}(t)-
\epsilon_{IJKL}\,E^{IJ}(t')\,E^{KL}(t')=0
\ee

\ni
and this, together with the simplicity constraint,
implies that the sum of the
two bivectors associated to the two triangles is again a simple bivector.
This implements the third of the Barrett-Crane constraints
(intersection constraint).

\mn
Let's note that in the Euclidean case, we can use the decomposition
into selfdual and anti-selfdual components to write the previous
constraints (for $t=t'$ or $t$ and $t'$ sharing an edge) as:

\be
\delta_{IJ}\left[
E^{(+)I}(t)E^{(+)J}(t')-E^{(-)I}(t)E^{(-)J}(t')\right]\,=\,0
\ee

\ni
thus showing that the simplicity for the bivectors $E(t)$ and $E(t)+E(t')$
is that their selfdual part and anti-self part have the same magnitude.
We note that in the Lorentzian case this decomposition implies
a complexification of the fields, so the physical interpretation is
somehow more problematic. However it presents no problems formally , and
corresponds to the splitting of the Lie algebra of $so(3,1)$, to which the
bivectors in Minkowski space are isomorphic, into $su(2)_{C}\oplus
su(2)_{C}$.

\mn
Thus, it is clear that the complicated constraints \Ref{whole} and
\Ref{simple2} for the field $B$ are just the Plebanski constraint for the
field $E$, associated to $B$ by means of the transformation \Ref{transf},
and, when discretized, are exactly the Barrett-Crane constraints. The
fields characterizing the 4-geometry of the triangulated manifold are then
the bivectors $E(t)$.

\mn
This result, by itself, does not imply necessarily that a spin foam
quantization of the generalized BF-type action gives the Barrett-Crane
model, but it means anyway that the geometry is still captured by a field
which, when discretized, gives a set of bivectors satisfying the
Barrett-Crane constraints. This in turn suggests strongly that the
Barrett-Crane constraints characterize the quantum geometry also in this
case, even if a first look at the constraints seemed to contradict this,
and consequently the simple representations are the right representations
of $so(4)$ and $so(3,1)$ that have to be used in constructing the spin
foam.   

\section{Spin foam quantization and constraints on the representations}
\label{quant}

\ni
We have already proven that the spin foam quantization cannot be performed
using a naive association between the $B$ field and the canonical
generators of the Lie algebra. On the other hand, we have seen that the
$B$ field can be put in correspondence with a field $E$ such that the
constraints on this are the Barrett-Crane constraints. This suggest that a
similar transformation between Lie algebra elements would make everything
work again, giving again the simplicity conditions for the
representations, as in the Barrett-Crane model.

\mn 
In light of the discussion in section \Ref{BC} on the natural way to
associate a bivector to a triangle in a gravitational context, using the
frame field, and also because the 2-forms $*(e^{I}\w e^{J})$ are a basis
for the space of 2-forms, we could argue that it is the bivector coming
from $*(e\w e)$ that has to be associated to the canonical generators of
the Lie algebra.    

\ni
That this is the right choice can be proven easily. In fact the
isomorphism between bivectors and Lie algebra elements is realized
choosing a basis for the bivectors such that they are represented by 4 by
4 antisymmetric matrices, and interpreting these matrices as being the
4-dimensional representation of Lie algebra elements. If this is done for
the basis 2-forms $*(e\w e)$, the resulting matrices give exactly the
canonical generators of $so(4)$ or $so(3,1)$, so that $*(e \w e)\,
\leftrightarrow\,J$.    

\ni
Then equation \Ref{B} suggests us that the field $B$ has to be associated
to elements $\tl{J}$ of the Lie algebra such that:

\be
B \leftrightarrow \tl{J}^{IJ}\,=\,\alpha\, J^{IJ}\,+\,\epsilon\,\beta
*J^{IJ}
\label{modif}
\ee

\ni
This is simply a change of basis (but not a Lorentz rotation),
since the transformation is invertible,
provided that $\alpha^{2}-\epsilon\beta^{2}\neq 0$. In some sense, we can
say that working with the $B$ in the action is like working with a
non-canonical basis in the Lie algebra, the canonical basis being instead
associated to $*(e\w e)$.

\mn
Now we consider the constraint \Ref{simple2} on the $B$ field, and use the
correspondence above to translate it into a constraint on the
representations of the Lie algebra.
The Casimir corresponding to $\f{1}{2}\epsilon_{IJKL}B_{ab}^{IJ}B^{abKL}$
is

\be
\tl{C}_2=\f{1}{2}\epsilon_{IJKL}\tl{J}^{IJ}\tl{J}^{KL}
=2\alpha \beta C_1 +(\alpha^2+\epsilon\beta^2)C_2
\ee

\ni
where $C_1$ and $C_2$ are the usual Casimirs associated to the canonical
generators $J^{IJ}$ (being $C_{1}=j^+(j^+ +1) + j^-(j^- +1)$ and
$C_{2}=j^+(j^+ +1) - j^-(j^- +1)$ in the Euclidean case, and $C_{1}=j^2
-\rho^2 -1$ and $C_{2}=\f{1}{2}j\rho$ in the Lorentzian case), while the
Casimir
associated to $B^{IJ}_{ab}B_{IJ}^{ab}$ is:

\be
\tl{C}_1=\tl{J}^{IJ}\tl{J}_{IJ}
=(\alpha^2+\epsilon\beta^2)C_1 + 2\epsilon\alpha \beta C_2
\ee

\ni
Substituting these expression in \Ref{simple2}, we get:

\be
2\alpha\beta \tl{C}_1=(\alpha^2+\epsilon\beta^2)\tl{C}_2\,\,
\Rightarrow\,\,
(\alpha^2-\epsilon\beta^2)^2\,C_2\,=\,0
\ee

\ni
In the assumed case $\alpha^2\ne \epsilon\beta^2$,
we find the usual Barrett-Crane
simplicity condition $C_2=0$ with a restriction to the simple
representations of $so(4)$ ($j^+=j^-$)
or $so(3,1)$ ($n=0$ or $\rho=0$). We see that, at least for what
concerns the representations to be used in the spin foam model, the whole
modification of the inital action
is absorbed by a suitable redefinition of the correspondence between
the field $B$ and the generators of the Lie algebra.

\mn
Now we want to discuss briefly how general is the association we used
between the $B$ field and Lie algebra elements, i.e. how many other
choices would give still the Barrett-Crane simplicity constraint on the
representations starting from the constraint \Ref{simple2} on the $B$
field.
Suppose we associate to $B$ a generic element $\tl{J}$ of the Lie algebra,
related to the canonical basis by a generic invertible transformation
$\tl{J}=\Omega J$. Any such transformation can be split into
$\tl{J}^{IJ}=\Omega^{IJ}_{KL}J^{KL}=(\alpha I+\epsilon\beta
*)^{IJ}_{MN}U^{MN}_{KL}J^{KL}=(\alpha I+\epsilon\beta
*)^{IJ}_{MN}J'^{MN}$, for $\alpha^2\ne \epsilon\beta^2$, so shifting all
the ambiguity into $U$, Inserting this into the constraints we obtain the
condition $C'_{2}=0$, where $C'_2=J'*J'$. If we now require that this
transformation should still give the Barrett-Crane simplicity constraint
$C_2=0$, then this amount to require $C'_2=\lambda C_2$ for a generic
$\lambda$. But if $C_2=0$ and the transformation preserves the second
Casimir modulo rescaling, then it should preserve, modulo rescaling, also
the first one. This means that the transformation $U$ preserves, modulo
rescaling, the two bilinear forms in the Lie algebra which the two
Casimirs are constructed with, i.e. the \lq\lq identity" and the
completely antisymmetric 4-tensor in the 6-dimensional space of
generators. Consider the Euclidean case. The set of transformations
preserving the first is given by $O(6)\times Z_{2}\simeq
SO(6)\times Z_{2}\times Z_{2}$, while the set of
transformations preserving the second is given by
$O(3,3)\times Z_{2}\simeq
SO(3,3)\times Z_{2}\times Z_{2}$, so that the $U$
preserving both are given by the intersection of the two groups, i.e. by
the transformations belonging to a common subgroup of them. Certainly a
common subgroup is given by $SO(3)\times
SO(3)\times Z_{2}\times Z_{2}\simeq
SO(4)\times Z_{2}\times Z_{2}$, and we can conjecture that
this is the largest one, since it covers all the symetries of the original
action, so that any solution of the theory, like \Ref{B} should be defined
up to such transformation, and we expect this to be true also in the
association between fields and Lie algebra elements. Of course we can in
addition rescale the $\tl{J}$ with an arbitrary real number. The argument
in the Lorentzian case goes similarly.

\mn
Coming back to the spin foam model corresponding to the new generalized
action, our results prove that it should still be based on the simple
representations of the Lie algebra, and that no ambiguity in the choice of
the labelling of the spin foam faces results from the more general form of
the constraints, since this can be naturally and unambiguously re-absorbed
in the correspondence between the $B$ field and the Lie algebra elements.
This suggests strongly that the resulting spin foam model corresponding to
the classical action here considered is still the Barrett-Crane model, as
for the Plebanski action, but it is not completely straightforward to
prove it due to the more complicated form of the action \Ref{after}.
Anyway a motivation for this is provided by the fact that our analysis
shows that the physics is still given by a set of bivectors $E$, in 1-1
correspondence with the field $B$ on which the generalized action is
based, and that this bivectors satisfy the Barrett-Crane constraints at
the classical level, with the translation of them at the quantum level
being straightforward. The action is in fact that of BF theory plus
constraints, which, as shown in this section, at the quantum level are
just the Barrett-Crane constraints on the representations used as
labelling in the spin foam, so that a spin foam quantization procedure of
the type performed in \cite{daniele} seems viable.

\section{Alternative: the Reisenberger model} \label{reis}

\ni
We have seen that a natural discretization and spin foam quantization of
the constraints in this generalized BF-type action leads to the
Barrett-Crane model. In this section we want to discuss and explore a bit
the alternatives to our procedure, and the cases not covered in our
previous analysis. 

\mn
In order to prove the equivalence of the two form of the constraints
\Ref{simple} and \Ref{simple2}, we assumed that $a_2\neq 0$ and that
$\left(\f{2a_2}{a_1}\right)\neq\epsilon$, which, in terms of $\alpha$ and
$\beta$ is requiring that $\alpha^2\neq\epsilon\beta^2$
($\alpha\neq\pm\beta$ in the Euclidean case, and $\alpha\neq\pm i\beta$ in
the Lorentzian one). Later, the transformations we used both at the
classical level ($B\rightarrow E$) and at the quantum (better, Lie
algebra) level ($\tl{J}\rightarrow J$) were well-defined, i.e. invertible,
provided that $\alpha^2\neq\epsilon\beta^2$, again.
Actually, the only interesting condition is the last one, since we already
mentioned at the beginning (section \Ref{action}) that the case $a_2=0$
leads to  considering only degenerate $B$ fields and degenerate
tetrads (in the Euclidean case). 

\mn
The case $\left(\f{2a_2}{a_1}\right)\neq\epsilon$ or
$\alpha^2\neq\epsilon\beta^2$ corresponds, at the canonical level, to the
Barbero's choice of the connection variable (with Immirzi parameter
$\gamma=\pm 1$) in the Euclidean case, and to Ashtekar variables
($\gamma=\pm i$) in the Lorentzian case. It amounts to formulate the
theory using a selfdual (or anti-selfdual) 2-form field $B$, or
equivalently a selfdual (or anti-self dual) connection. In fact, a look at
the constraints \Ref{simple} shows clearly that, in this case, they
exactly imply the selfduality (or anti-selfduality) of the field $B$.
There is no rigorous way to relate the Barrett-Crane conditions and spin
foam model
to the classical action when this happens, at least using our procedure,
since the equivalence of \Ref{simple} and \Ref{simple2} cannot
be proved and the transformations we used are not invertible.
The constraints on the field $B$ just
state that $B$ is the (anti-)selfdual part of a field $E$ (and
in this case, the action \Ref{action} corresponds to the (anti-)selfdual
Plebanski action for $E$).
We note however that if we use still the constraints in the form
\Ref{simple2} in the Euclidean case, in spite of the fact that we are not
able to prove their equivalence with the original ones, and translate them
into a constraint on the representations of $so(4)$ using the naive
correspondence $B\rightarrow J$, where $J$ is the canonical basis of the
algebra, we get: $C_1=\pm C_2$ for $2a_2=\pm a_1$. The first case leads to
$j^-=0$ and the second to $j^+=0$, so we are reduced from $so(4)$ to
$su(2)_L$ or to $su(2)_R$, with a precise correspondence between the
(anti-)selfduality of the variables used and the (anti-)selfduality of the
representations labelling the spin foam.
There exist a spin foam model in these ``degenerate'' cases.
It is the Reisenberger
model for left-handed (or right-handed) Euclidean gravity
\cite{mike}\cite{mike2}, whose relationship with the Barett-Crane model is
unfortunately not yet clear.
This model \cite{mike}\cite{mike2} can be associated to a different
discretization of the generalized action \Ref{action} using the original
form
\Ref{simple} of the constraints, as it was
analyzed in \cite{dpf} for the Plebanski action.
This is indeed the only other (known)
alternative to our procedure, at least in
the Euclidean case. 

\ni
As before we define the volume spanned by the
two triangles $S$ and $S'$:

\be
V(S,S')=\int_{x\in S, y\in S'}e\,\epsilon_{abcd}\,
dx^a\w dx^b\w dy^c\w dy^d
\ee

\ni
To use \Ref{simple}, we decompose the 2-form B inside
the 4-simplex into a sum of 2-forms associated to the faces (triangles)
of the 4-simplex \cite{dpf}\cite{mike}\cite{F-K}:

\be
B^{IJ}(x)=\sum_S B^{IJ}_S(x)
\ee

\ni
where $B^{IJ}_S(x)$ is such that

\be
\int B^{IJ}_S\w J=B^{IJ}[S]\int_{S*}J
\ee

\ni
with $J$ is any 2-form and $S*$ the dual face of $S$ (more precisely
the wedge dual to $S$, i.e the part of the dual face to $S$
lying inside the considered 4-simplex).
Then, it is clear that:

\bean
\int_S B^{IJ}_{S'}=\delta_{S,S'}B^{IJ}[S] \\
\int B^{IJ}_S\w B^{KL}_{S'}=B^{IJ}[S]B^{KL}[S']\epsilon(S,S')
\eean

\ni
where $\epsilon(S,S')$ is the sign of the oriented volume $V(S,S')$.
More precisely $\epsilon(S,S')=\pm 1$ if $S$, $S'$ don't share any edge,
and $\epsilon(S,S')=0$ if they do. 

Using that, we can translate \Ref{whole} and \Ref{simple} into:

\be
\tl{\Omega}^{IJKL}=\Omega^{IJKL}-
\f{1}{6}\eta^{[IK}\eta^{J]L}\Omega^{AB}_{AB}-
\f{1}{24}\epsilon^{IJKL}\epsilon_{ABCD}\Omega^{ABCD}=0
\ee

and 
\be
4a_2\Omega^{AB}_{AB}=a_1\epsilon_{ABCD}\Omega^{ABCD}
\ee

\ni
where

\be
\Omega^{IJKL}=\sum_{S,S'}B^{IJ}[S]B^{KL}[S']\epsilon(S,S')
\label{defomega}
\ee

\ni
These are the $so(4)$ analogs of the Reisenberger constraints.
Let's note that the constraints involve associations triangles
not sharing any edge whereas the Barrett-Crane procedure
was to precisely study triangles sharing an edge i.e being in
the same tetrahedron. This is one of the reasons why it is hard to link
these two models.
Then following \cite{mike2}\cite{F-K}, it is possible to calculate
the amplitude associated to the 4-simplex and the corresponding spin foam
model. 
For this purpose, it is useful to project these constraint on the selfdual
and anti-selfdual sectors as in \cite{dpf}:

\be
\tl{\Omega}^{ij}_{++}=
\Omega^{ij}_{++}-\delta^{ij}\f{1}{3}tr(\Omega_{++})
\ee

\be
\tl{\Omega}^{ij}_{--}=
\Omega^{ij}_{--}-\delta^{ij}\f{1}{3}tr(\Omega_{--})
\ee

\be
\Omega^{ij}_{+-}=0
\ee

\be
\Omega_0=
(\alpha-\beta)^2tr(\Omega_{++})+
(\alpha+\beta)^2tr(\Omega_{--})=0
\ee

\ni
The two first constraints are the same as in the Reisenberger model
for $su(2)$ variables. The two last constraints link the two sectors
(+ and -) of the theory, mainly stating that they is no correlation
between them except for the constraint $\Omega=0$.
Indeed,  only that last constraint is modified by the introduction
the Immirzi parameter. Following the notations of \cite{mike2},
we replace the field $B^{IJ}$ by the generator $J^{IJ}$ of $so(4)$
(this is a formal quantization of the discretised BF action, see
\cite{mike2}
for more details) and we define the projectors $P_{1,2,3,4}$
(or some gaussian-regularised projectors) on the
kernel of the operators corresponding to the four above constraints.
Then, the amplitude for the vertex $\nu$ is
a function of
the holonomies on the 1-dual skeleton of the three-dimensional
frontier $\pp\nu$ of the vertex. It is given by
projecting an universal state (or topological state since that without
the projectors, the amplitude gives the $so(4)$ Ooguri's topological
model)
and then integrating it over
the holonomies around the ten wedges $\{h_l\}_{l=1\dots10}$:

\be
a(g_{\pp\nu})=\int dh_l \prod_{s\,\tr{\tiny wedge}}
\sum_{j_s}tr^{j^1\otimes\dots\otimes j^{10}}\left[
P_1P_2P_3P_4\bigotimes_s(2j_s+1)U^{(j_s)}(g_{\pp s})
\right]
\ee

\ni
It seems that a vertex including the Immirzi parameter is
perfectly well-defined in this case. Let's analyse this
more closely.
As only one constraint involves the Immirzi parameter,
in a first time, we will limit ourselves to studying its action.
Replacing the field $B$ by the generators $J^{IJ}$, we get

\bea
\Omega_0&=&
\sum_{S,S'}\epsilon(S,S')\left[
(\alpha-\beta)^2
J^{+ i}[S]J^{+ i}[S']
+(\alpha+\beta)^2 (\tr{antiselfdual})
\right]
\nonumber  \\
&=&
\sum_{S,S'}\epsilon(S,S')\f{1}{2}\left[
(\alpha-\beta)^2
\left((J^{+ i}[S]+J^{+ i}[S'])^2-J^{+ i}[S]^2-J^{+ i}[S']^2\right)
+(\alpha+\beta)^2 (\tr{antiselfdual})
\right]
\eea

\ni
So $\Omega_0$ can be expressed in term of the ($su(2)$) Casimir
of the representations associated to each wedge $S$ and
the Casimir of their tensor products.
The difference from the Barrett-Crane
model is that it is the tensor product of the representations
associated to two triangles which do {\it not} belong to the same
tetrahedron, so there is little hope of finding directly
an equivalent of the simplicity/intersection constraints.
Nevertheless, we can do a naive analysis.
Casimirs will always give numbers $j(j+1)$. But we need to choose a basis
of $j^1\otimes\dots\otimes j^{10}$. So calculating the action of
$\Omega_0$ might involve some change of basis and thus some Clebsch-Gordon
coefficient. However, those are still rational. 
So we conjecture that the amplitude of the vertices will be zero
(no state satisfying $\Omega_0=0$)
except if $\alpha=\beta$, $\alpha=0$ or $\beta=0$ as in our
first approach to quantizing the $B$-field constraints in
the Barrett-Crane framework.
If this conjecture is verified, we will have two possibilities:
either the discretization procedure
is correct and we are restricted to a few consistent cases, or we need to
modify the discretization or quantization procedures.

\mn
Resuming, the situation looks as follows. We have the most general BF-type
action for gravity, depending on two arbitrary parameters, both in the
Euclidean and Lorentzian signature. In both signatures, and for all the
values of the parameters except one (corresponding to the Ashtekar-Barbero
choice of canonical variables), the constraints which give gravity from BF
theory can be expressed in such a form that a spin foam quantization of
the theory leads to the Barrett-Crane spin foam model. In the Euclidean
case, for all the values of the parameters, a different discretization,
and quantization procedure, leads to the Reisenberger model, but it also
seems that the last spin foam model is non-trivial (i.e. non-zero vertex
amplitude) only for some particular values of the parameters $\alpha$ and
$\beta$. These two spin foam models may well turn out to be equivalent,
but they are a priori
different, and their relationship is not known at this stage. There is no
need to stress that an analysis of this relationship would be of paramount
importance.
   
\section{The role of the Immirzi parameter in the spin foam models}
\label{area}

\ni
We would like to discuss briefly what our results suggest regarding the
role of the Immirzi parameter in the spin foam models, stressing that this
suggestion can at present neither be well supported nor disproven by
precise calculations.
As we said in section \Ref{CMPR}, the BF-type action \Ref{action}, after
the imposition of the constraints on the field $B$, reduces to a
generalized Hilbert-Palatini action for gravity, in a form studied within
the canonical approach in \cite{holst}. The canonical analysis performed
in that work showed that this action is the lagrangian counterpart of the
Barbero's hamiltonian formulation \cite{barbero} introducing the Immirzi
parameter \cite{immirzi} in the definition of the connection variable and
then in the area spectrum. This led to the suggestion \cite{CMPR} that
the spin foam model corresponding to the new action \Ref{action} would
present non-simple representations and an arbitrary (Immirzi) parameter as
well.  

\mn 
On the contrary we have shown that the spin foam model corresponding to
the new action is given again by the Barrett-Crane model, with the
representations labelling the faces of the 2-complex being still the
simple representations of $so(4)$ or $so(3,1)$. The value of the area of
the triangles in this model is naturally given by the (square root of the)
first Casimir of the gauge group in the representation assigned to the
triangle, with no additional (Immirzi) parameter. From this point of view
it can be said that the prediction about the area spectrum of spin foam
models and canonical (loop) quantum gravity do not coincide. 

\mn
However both the construction of the area operator and its diagonalization
imply working with an Hilbert space of states, and not with their
histories as in the spin foam context, and the canonical structure of the
spin foam models, like the Barrett-Crane one, is not fully understood yet.
Consequently, the comparison with the
loop quantum gravity approach and results is not straightforward. In fact,
considering for example the Barrett-Crane model, it assigns an Hilbert
space to boundaries of spacetime, and these correspond, in turn, to
boundaries of the spin foam, i.e. spin networks, so that again a spin
network basis spans the Hilbert space of the theory, like in loop quantum
gravity. The crucial difference, however, is that the spin networks  used
in the Barrett-Crane model are constructed out of (simple) representation
of $so(4)$ or $so(3,1)$, i.e. the full local gauge group of the theory,
while in loop quantum gravity
(for a nice introduction see \cite{rovelli&gaul})
the connection variable used is an
$su(2)$-valued connection resulting from a breaking of the gauge group
from $so(4)$ or $so(3,1)$ to that subgroup, in the process of the
canonical $3+1$ decomposition, so that the spin network basis uses only
$su(2)$ representations. Consequently, a comparison of the results in spin
foam models could possibly be made more easily
with a covariant (with respect to the gauge
group used) version of loop quantum gravity, i.e. one in which the group
used is the full $so(4)$ or $so(3,1)$.

\mn
The only results in this sense we are aware of were presented recently in
\cite{alex}\cite{alex2}.
In these two papers, a Lorentz covariant version of loop
quantum gravity is sketched at the algebraic level. However, the
quantization was not yet achieved and 
the Hilbert space of the theory (``spin networks'') not
constructed because of problems arising from the non-commutativity of the
connection variable used.
Nevertheless, two results were derived from the formalism.
The first one is that the path integral of the theory
formulated in the covariant variables is independent 
from the Immirzi parameter \cite{alex}, which becomes
an unphysical parameter whose role is to regularize the theory.
The second result was the construction of an area operator acting
on the hypothetic ``spin network'' states of the theory \cite{alex2}:

\be
{\cal A} \approx   l_P^2\sqrt{-C(so(3,1))+C(su(2))}
\label{spectrum}
\ee

\ni
where $C(so(3,1))$ is a quadratic Casimir of the Lie algebra $so(3,1)$
(the Casimir $C_1$ to a factor)
and $C(su(2))$ the Casimir of the spatial pull-back $so(3)$. So a
``spin network'' state would be labelled by both
a representation of $so(3,1)$
and a representation of $su(2)$. As we see, the area, whose spectrum
differ from both the the spin foam and the loop quantum gravity one, is
independent
on the Immirzi parameter. However, we do not know yet if such an area
spectrum has any physical meaning, since no Hilbert space has been
constructed for the theory.
Nevertheless, it suggests that a ``canonical'' interpretation
of a spin foam might not be as straightforward as it is believed. Instead of
taking as spatial slice a $SO(3,1)$ spin networks by cutting a spin foam
\cite{gns}, we might have to also project the $SO(3,1)$ structure onto
a $SU(2)$ one; the resulting $SU(2)$ spin network been our space and
the background $SO(3,1)$ structure describing its space-time embedding.

\mn
It is clear that this issue has not found at present any definite
solution, and it remains rather intricate. 
Thus all we can say is that our results (that do not regard directly the
issue of the area spectrum) and those in
\cite{alex2} suggest that the appearence of the Immirzi parameter in
loop quantum gravity is an indication of the presence of a quantum
anomaly, as discussed in \cite{tom&carlo}, but not of a fundamental one,
i.e. not one originating from the breaking up of a classical symmetry at
the quantum level, and indicating that some new physics takes place.
Instead what seems to happen is that the symmetry is broken by a
particular choice of quantization procedure, and that a fully covariant
quantization, like the spin foam quantization or the manifestly Lorentzian
canonical one, does not originate any one-parameter ambiguity in the
physical quantities to be measured, i.e. no Immirzi parameter. Of course,
much more work is needed to understand better this issue, in particular
the whole topic of the relationship between the canonical loop quantum
gravity approach and the covariant spin foam one is to be explored in
details, and to support or disprove this idea.

\section{Conclusions}
To conclude, we have shown that starting from the most general BF-type
action for gravity we are naturally lead to the Barrett-Crane spin foam
model as its corresponding quantization. More precisely, the $B$ field in
that action can be put in 1-1 correspondence with a set of bivectors such
that the constraints on them are exactly the Barrett-Crane ones, at the
classical level; moreover, a translation of these at the quantum level,
based on the association between $B$ field and Lie algebra elements, gives
exactly the simplicity constraint on the representations of $so(4)$ or
$so(3,1)$ to be used in the spin foam model. We also explored the possible
alternatives, all leading to the Reisenberger model in the Euclidean case,
based on (anti-)selfdual fields at the classical level and on the
$su(2)_L$ ($su(2)_R$) subset of representations of $so(4)$ at the quantum
one. Regarding the possible role of the Immirzi parameter in the spin foam
models, our results suggest that it does not appear in the quantization of
the generalized action, in agreement with the idea that its appearence is
a result of the breaking of Lorentz covariance in the usual canonical
approach. This issue, however, remains to be understood. Further work is
also needed to investigate the relationship between
the Reisenberger model and the Barrett-Crane one, in order to understand
whether they represent two formulation of the same quantum theory or two
different and inequivalent quantum models. In light of the results
presented here, we believe this is really a central issue.  

\vspace{-2mm}

\section*{Acknowledgments}

\ni
R.L. would like to thank S. Alexandrov for his useful explanations
on the covariant canonical formalism,
while D.O. would like to thank H. Pfeiffer
for many useful discussions and suggestions on group and representation
theory. 

\vspace{-2mm}

\appendix

\section{Equivalence of the two forms of constraints}

\ni
Here, we follow the same procedure as in \cite{dpf} to show the
equivalence of the two forms of the constraints for the $B$ field,
the one leading to the disrete constraints as needed in the
Reisenberger model and the one leading to the Barrett-Crane discrete
geometry picture. First, the constraints \Ref{mixed} and \Ref{simple}
can be condensed into a single equation:

\be
\epsilon^{abcd}B^{IJ}_{ab}B^{KL}_{cd}=
\epsilon\,e\left(
\epsilon^{IJKL}+\f{a_1}{a_2}\eta^{[I\mid K\mid}\eta^{J]L}
\right)
=e\Omega^{IJKL}
\label{app1}
\ee

\ni
Considering all the variables as $6\times 6$ matrices
(on the antisymmetric couples $[IJ]$ and $[ab]$), we write
the previous equation as:

\be
B_{IJ}^{ab}\epsilon_{ab}^{cd}B_{cd}^{KL}=e\Omega_{IJ}^{KL}
\label{app2}
\ee

\ni
Then $\Omega_{IJ}^{KL}$ is invertible when

\be
\left(\f{a_1}{2a_2}\right)^2\ne \epsilon
\ee

\ni
Assuming that this is the case, and that
$e\ne0$ ($B$ field non-degenerate),
we can define

\be
\Sigma^{ab}_{IJ}=
\f{1}{e}\Omega_{IJ}^{KL}B_{KL}^{cd}\epsilon_{cd}^{ab}
=\f{\epsilon}{4e}\epsilon^{abcd}
\f{1}{\left(\f{a_1}{2a_2}\right)^2-\epsilon}
\left(
-\epsilon_{IJKL}+\f{a_1}{a_2}\eta_{[IK}\eta_{J]L}
\right)
B^{KL}_{cd}
\label{sigma}
\ee

\ni
and rewrite \Ref{app2} as:

\be
\Sigma^{cd}_{IJ}B^{KL}_{cd}=\delta^{KL}_{IJ}
\label{app3}
\ee

\ni
\Ref{app3} means that 
$\Sigma_{IJ}^{cd}$ and $B_{cd}^{KL}$
are invertible and
inverse of each other, so that it is equivalent to:

\be
B_{cd}^{IJ}\Sigma_{IJ}^{ab}=\delta_{cd}^{ab}
\label{app4}
\ee

\ni
Expanding this last equation and inverting $\epsilon^{abcd}$, we get:

\be
\left(
-\epsilon_{IJMN}+\f{a_1}{a_2}\eta_{[IM}\eta_{J]N}
\right)
B^{MN}_{cd}B^{IJ}_{ab}=
\epsilon\,e\,\epsilon_{abcd}
\left(\left(\f{a_1}{2a_2}\right)^2-\epsilon\right)
\ee

\ni
We can check that in the case $a_1=0$ (Plebanski action),
we find the same constraint as in \cite{dpf} which leads to
the Barrett-Crane simplicity constraint after discretization.

\mn
Let's note that in the Lorentzian case ($\epsilon=-1$),
the condition on $a_1,a_2$ is automatically satisfied if we keep real
variables.

\section{Selfdual and anti-selfdual components of the action}

\mn
In the Euclidean case, decomposing the action into
selfdual and anti-self dual components gives another way 
of understanding its structure.
Following \cite{dpf}, we decompose
$B=B^{(+)}+B^{(-)}$, $\omega=\omega^{(+)}+\omega^{(-)}$
and $\phi=\phi^{(+)}+\psi+\phi^{(-)}+\phi_0$. $\phi^{(+)}_{ij}$
and $\phi^{(-)}_{ij}$ are two symmetric traceless matrices (left and right
parts of the Weyl), $\psi_{ij}$ is an antisymmetric matrix (traceless
part of the Ricci) and $\phi_0$ is a scalar (the only remaining one
after imposing the constraint $H(\phi)=0$ on the field $\phi$.
The action \Ref{action} then writes:

\be
S=S^++S^-+S^{\psi}+S^0
\ee

\be
S^{\pm}=\int \left[\delta_{ij}B^{(\pm)i}\w F^{(\pm)j}
-\f{1}{2}\phi_{ij}^{(\pm)}B^{(\pm)i}\w B^{(\pm)j}
\right]
\ee

\be
S^{\psi}=\int \left[
-\psi_{ij}B^{(+)i}\w B^{(-)j}\right]
\ee

\be
S^0=\int\left[-\f{\phi_0}{2}\delta_{ij}
\left(
2a_2(B^{(+)i}\w B^{(+)j}+B^{(-)i}\w B^{(-)j})
-a_1(B^{(+)i}\w B^{(+)j}-B^{(-)i}\w B^{(-)j})
\right)
\right]
\label{scalar}
\ee

\ni
In this last equation \Ref{scalar}, we can also replace $2a_2$ by
$\alpha^2+\beta^2$ and $a_1$ by $2\alpha\beta$ using \Ref{a1a2}
and rewrite it as

\be
S^0=\int\left[-\f{\phi_0}{2}\delta_{ij}
\left(
(\alpha-\beta)^2B^{(+)i}\w B^{(+)j} +
(\alpha+\beta)^2B^{(-)i}\w B^{(-)j}
\right)
\right]
\ee

\ni
This motivates to
renormalise the selfdual and
anti-selfdual parts of $B$:

\be
B^{(\pm)}=(\alpha+\beta)E^{(\pm)}
\ee

\ni
And renormalising also the components of $\phi$ to absorb the changes,
we find that only the dynamical terms $B^{(\pm)}\w F^{(\pm)}$ get modified
and also the scalar constraint:

\be
B^{(\pm)}\w F^{(\pm)} = (\alpha \pm \beta)E^{(\pm)}\w F^{(\pm)}
\ee

\be
S^0=\int\left[-\f{\tl{\phi}_0}{2}\delta_{ij}
\left(
E^{(+)i}\w E^{(+)j} +E^{(-)i}\w E^{(-)j}
\right)
\right]
\ee

\ni
Thus we recover exactly the action \Ref{after}.

\end{document}